\documentclass[preprint,12pt,authoryear]{elsarticle}

\usepackage{amssymb,amsfonts}
\usepackage{graphicx}
\usepackage{bm}
\usepackage[retainorgcmds]{IEEEtrantools}
\usepackage{amsmath}

\journal{arxiv.org}

\begin{document}

\begin{frontmatter}
\title{Chaperone driven polymer translocation through Nanopore: spatial distribution and binding energy}
\author{Rouhollah Haji Abdolvahab}
\address{Physics Department, Iran University of Science and Technology (IUST), 16846-13114, Tehran, Iran.}
\ead{rabdolvahab@gmail.com}
\begin{abstract}
Chaperones are binding proteins which work as a driving force to bias the biopolymer translocation by binding to it near the pore and preventing its backsliding. Chaperones may have different spatial distribution. Recently we show the importance of their spatial distribution in translocation and how it effects on sequence dependency of the translocation time. Here we focus on homopolymers and exponential distribution. As a result of the exponential distribution of chaperones, energy dependency of the translocation time will changed and one see a minimum in translocation time versus effective energy curve. The same trend can be seen in scaling exponent of time versus polymer length, $\beta$ ($T\sim\beta$). Interestingly in some special cases \textit{e.g.} chaperones of size $\lambda=6$ and with exponential distribution rate of $\alpha=5$, the minimum reaches even to amount of less than $1$ ($\beta<1$). We explain the possibility of this rare result and base on a theoretical discussion we show that by taking into account the velocity dependency of the translocation on polymer length, one could truly predict the amount of this minimum.
\end{abstract}
\end{frontmatter}

\section{Introduction}
\label{intro}

Translocation of biopolymers through nanopores \cite{Meller03} is of fundamental importance in both biology and biotechnology. Cell metabolism contains vital instances of the process. Translocation of mRNA through nuclear pores after transcription is an example \cite{Al02}. Proteins translocation through endoplasmic reticulum or organelles like mitochondria are also some other important instances \cite{Al02,muthuAnn07,rapaport}. Drug delivery, gene therapy and fast cheap sequencing are some of its significant applications in biotechnology \cite{Marzio03,Nakane03,Branton08,Ramin12,Fanzio12,Wanunu15,Liang15}. After the study of Bezrukov \textit{et al.} on counting polyethylene oxide molecules using an alamethicin ion channel \cite{Bezrukov94nature} and seminal experimental work of Kasianowicz \textit{et al.} \cite{Kasianowicz96} on ssRNA translocation through an $\alpha$-hemolysin channel embedded in a bilayer, there has been flurry of theoretical works, experiments and simulations in the field \cite{Meller03,Panja13,Sun14,metzr14}.

Procedures which used \textit{in vivo} and \textit{in vitro} for driving the translocation are different. Although \textit{in vitro}, people use strong electric field as a driving force to pull the negatively charged biopolymers, there is not such a strong electric field \textit{in vivo} and cell selects binding proteins for driving the translocation \cite{Tomkiewicz07}. The model was first proposed by Simon \textit{et al.} in 1992 as Brownian ratchet mechanism \cite{Si&Pe&Os}. In this model some proteins, called chaperones, bind to the polymer and as a result of their size prevent polymer from backsliding and bias its translocation. Experimental work of Matlack \textit{et al.} in 1999 drew attention to the problem again \cite{L&R&H,elston02,Z&R}. In spite of the many works in theoretical aspects and simulations of chaperone driven polymer translocation, we has not reached to a universal consensus on its statistical parameters \cite{metzr14,A&M4,Abd08,AbdPRE11,AbdJCP11,kaifujacs,kaifuPRE14,Cao15,Suhonen16,AbdPLA16}.

The process is quite complex and there are many important factors in determining translocation time of the polymer. In our recent paper we introduced a new factor, spatial distribution of the chaperones, into account \cite{AbdPLA16} which our simulation results show importance of the factor in translocation time itself and its dependency on sequence. There could be different mechanisms creating the exponential distribution of chaperones near the membrane. The cell may localize protein synthesis. Transporting the synthesized Chaperones may also be another mechanism~\cite{Besse08,Wang07}, physically the membrane affinity itself could also help.

Here we focus on its effects on the dependency of translocation time on binding energy and demonstrate some interesting results and how to interpret them. We briefly describe our model and simulation and introduce the chaperone spatial distribution in the following section. We go through our results in section \ref{result} in detail and compare simulation result by speculated theory. Finally in section \ref{conc} we will draw our conclusion.

\section{Model and method}
\label{mm}

\subsection{Introducing model}
\label{model}

\begin{figure}
\includegraphics[width=0.75\textwidth]{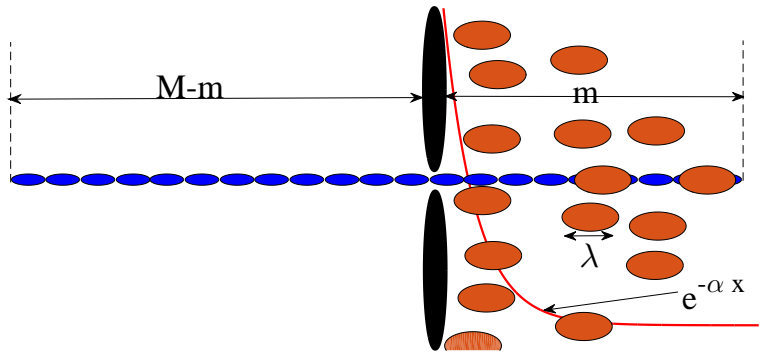}
\centering
\caption{Translocating 1d polymer, constructed of monomers of size
$\sigma$. Chaperones are only in the trans (right) side. They have the same size of
$\lambda\sigma$ in which $\lambda$ is an integer. The number $m$ of already
translocated monomers is a natural ``reaction coordinate'' of the
translocation process. The exponential curve ($\exp^{\alpha x}$) in the right part, shows the spatial distribution of chaperones.}
\label{Model}
\end{figure}

We have a stiff polymer of size $L=M\sigma$ in which $M$ and $\sigma$ are total number of the monomers and size of a single monomer, respectively. Chaperones, binding proteins, exist only in the \textit{trans}(right) side and have the same size of $\lambda\sigma$ in which $\lambda$ supposed to be an integer \cite{AbdPRE11,A&M4}(see the figure \ref{Model}). The wall has not any width and its average effect on polymer translocation is contained in polymer diffusion constant \cite{Cao11,Cao12}. Chaperones will bind (unbind) to (from) the polymer continuously in each step of the polymer translocation. Their size are greater than the pore and therefore the polymer could not come back when a chaperone is bound to it near the wall. Consequently its movement biased through the \textit{trans} side. Due to its length and interaction by the pore walls, the polymer has an effective diffusion which is quite small with regards of chaperones \cite{A&M4}. Consequently one can mean over the number of chaperones bound to the polymer and approximate the master equation for this problem as a $1+1$ dimensional equation:

\begin{eqnarray}
\label{me}
\frac{\partial P(m,t)}{\partial t} &=& W^+(m-1)P(m-1,t-1)\cr
\nonumber \\&+&W^-(m+1)P(m+1,t+1)\cr
\nonumber \\&-&(W^+(m)+W^-(m))P(m,t)
\end{eqnarray}

where $W^\pm(m)$ are the transfer rates of translocating polymer to the right and left when its location is $x=m\sigma$. Knowing those rates and the boundary conditions, here reflecting at first and absorbing at the end, one could calculates the mean passage time of the polymer translocation \cite{Ga}.

\textit{Important units:} Here we use of $\sigma$, Size of a monomer, as the unit of length. Energy is characterized by unit of $k_BT$ for simplicity. $\tau_0$, the typical time it takes for uncovered polymer to translocate over the distance $\sigma$ is used as our time unit. Hereafter everywhere we use the terms length or size, energy and time without pointing out their units the units are $\sigma$, $k_BT$ and $\tau_0$ respectively.

\subsection{Simulation}
\label{sim}
Translocation of a stiff polymer through nanopore using a dynamical Monte Carlo has been simulated. We apply the famous transmission boundary condition \cite{redner}, reflective at first and absorbing at the end, in calculating the translocation time. By this conditions we mean the first monomer of the polymer is always in the \textit{trans} side and when the whole polymer translocate through the \textit{trans} side, it could not come back again. Polymer goes to the right(\textit{trans}) side with probability of $1/2$. As a result of the chaperone's binding to the polymer, backsliding through \textit{cis} is not always possible. The backsliding is only feasible in case that there is not any chaperones bound to the polymer near the wall. In that situation the polymer goes to the right or remain at its place with the same probability.

In each step of polymer translocation the chaperones will try to bind/unbind $5m$ times, in which $m$ is the number of monomers translocated to the right side \cite{A&M4,AbdPRE11}. Tree terms are involved in chaperones binding probability: Boltzmann distribution which considered the binding energy between monomers and the chaperone, entropy related to the different binding patterns of the chaperones and the loss of entropy due to the decreasing of the chaperones number available in the space \cite{A&M4}. The second term is already exist in simulation process and the first and the last term is considered through an effective binding energy, $E_{eff}$ which takes into account the effect of chaperones concentration \cite{AbdPRE11}:

\begin{equation}
\label{ebe}
E_{eff}^i\equiv-\frac{1}{\lambda}\log\left[c_0v_0\exp\left(-
\frac{\varepsilon_i}{k_BT}\right)\right].
\end{equation}
Here $\varepsilon_i$ denotes the chaperone binding energy per monomer of the polymer, $c_0$ stands for the chaperone concentration, and $v_0$ is their volume \cite{AbdJCP11}. Therefore the binding and unbinding probability is write as:

\begin{eqnarray}
\nonumber
P_{bind}   =\frac{\exp\left(-\sum_{i=1}^\lambda E_{eff}^i\right)}{1+\exp\left(-\sum_{i=1}^\lambda E_{eff}^i\right)},
P_{unbind} = \frac{1}{1+\exp\left(-\sum_{i=1}^\lambda E_{eff}^i\right)}.
\label{pr}
\end{eqnarray}

Translocation process is repeated for at least $10^4$ times to obtain an acceptable error in reporting the mean first passage time. Size of the chaperones, $\lambda$, is changed from $1$ to $10$. To calculate the scaling exponents we also change the polymer length from $50$ to $300$ monomers.

\subsection{Exponential spatial distribution for chaperones}
\label{dist}

Binding particles, chaperones, may have different spatial distributions. In addition to uniform distribution in which presume in literature \cite{Abd08,AbdPRE11,AbdJCP11,A&L&M,A&M4}, recently we showed the important effects of the exponential distribution of chaperones in sequence dependency of the polymer \cite{AbdPLA16}. It is important here to note that due to binding and unbinding of chaperones in a dynamical situation, cell should consume energy to maintain an exponential distribution.

 Auspiciously, definition of the $E_{eff}$ allow us to easily take the spatial distribution into account in our simulation. One should just change the chaperone's effective binding energy from $E_{eff}$ to $E_{eff} + \alpha n$, where $\alpha$ is the exponential rate and $n$ is the distance between wall and the monomer in which we try to bind a chaperone (\textit{cf.} equation \ref{ebe}). Consequently chaperone's concentration will change from $c_0$ to $c(n)=c_0\exp\left(\lambda(-\alpha n)\right)$ \cite{AbdPLA16}.

\section{Results and discussions}
\label{result}

\subsection{Theoretical expectations}
\label{theoexp}
In order to bind a chaperone with size of $\lambda$, there should be a free space of the same size near the wall, to bias the translocation through the \textit{trans} side. Probability of binding a chaperone of size $\lambda=1$ near the wall in equilibrium can be calculated by its Boltzmann distribution as:

\begin{equation}\label{Pnweq}
P^{\lambda=1,Eq}_{nw}(E_{eff})=(\frac{\exp^{- E_{eff}}}{1+\exp^{- E_{eff}}})
\end{equation}

which does not depend on exponential rate $\alpha$. However, in the case of $\lambda>1$, one should consider the whole partition function \cite{AbdPLA16} to calculate the equilibrium probability. Non-equilibrium behaviour in our simulation results and the complexity of translocation of a finite polymer led us to consider the dynamic of the problem in our estimation. To take into account the dynamics of the problem, we pay our attention to the time in which a chaperone stay in its position or the position remain bare of them. Those times can be estimated using the Boltzmann distribution as follows. Assume a chaperone is bound to the polymer near the wall. The time in which the chaperone stay in its place, its dwell time, should be proportional to $P^\lambda_0=\frac{\exp^{-\lambda E_{eff}}}{1+\exp^{-\lambda E_{eff}}}$. We estimate the time in which the chaperone are bound to the polymer with one monomer distance from the wall as $P^\lambda_1=\frac{\exp^{-\lambda E_{eff}-\alpha}}{1+\exp^{-\lambda E_{eff}-\alpha}}$. Continuing this procedure led us to the following estimation for $P^{\lambda}_{nw}$:

\begin{equation}\label{Pnwapp}
P^{\lambda}_{nw}(\alpha,E_{eff})=(\frac{P^\lambda_0}{\sum^{\lambda}_{i=0} (P^\lambda_i)})
\end{equation}

where $P^\lambda_i$ is defined as:

\begin{eqnarray}\label{Pnwapp1}
P^\lambda_i&\equiv&(\frac{\exp^{-\lambda E_{eff}-i\alpha}}{1+\exp^{-\lambda E_{eff}-i\alpha}}), (i=0,1,\cdots,\lambda-1),\cr
P^{*\lambda}_{i=\lambda}&\equiv&(\frac{1}{1+\exp^{-\lambda E_{eff}}}).
\end{eqnarray}

$P^{*\lambda}_{i=\lambda}$ corresponds to the times in which $\lambda$ monomers near the wall are unbounded regardless of other monomers to be bounded or unbounded. Considering large positive values of $E_{eff}$ where chaperones dislike to bind, it goes to zero and in limits of large negative $E_{eff}$, high affinity of chaperones, it reaches to $1/\lambda$ (corresponds to the minimum time of $\lambda M$ in appendix \ref{app21}). Pay attention also that in spacial case of $\lambda=1$ we will reach to the Boltzmann distribution; equation \ref{Pnweq}. The passage time calculated by this result predict the behavior but its minimum location is not exact, \textit{e.g.} in the case of $\lambda=2$ minimum of the $E_{eff}$ becomes $-\frac{1}{2}\ln(\exp^{\alpha}(1+\exp^{2\alpha}\sqrt{\exp^{\alpha}-1})/(\exp^{\alpha}-2))$ which for $\alpha=5$ becomes $1.295$. For the case of $\lambda=6$ and $\alpha=5$ one also could find it numerically as $-0.431$ which differs from simulation results(for $\lambda=2, \alpha=5$ is $-2$ and for  $\lambda=6, \alpha=5$ nearly equals to $-1$; Fig.\ref{TEn}).

The scaling exponent $\beta$ also can be calculated within the context of convection-diffusion. Taking into account the length dependency of translocation velocity led us to quite good approximation of the minimum exponent location (see the appendix \ref{app3} and the discussion in its last paragraph.).

\subsection{Binding energy and chaperone's size}
\label{enl}
\begin{figure}
\includegraphics[width=10cm]{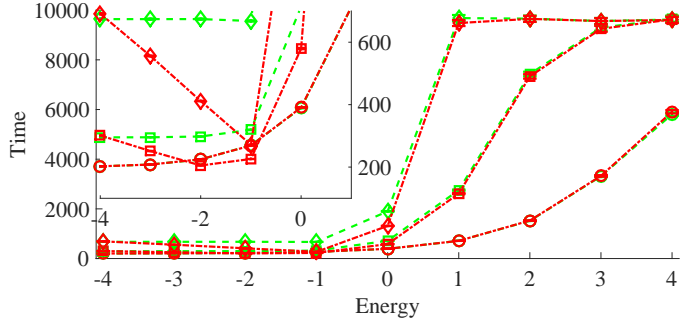}
\centering
\caption{Mean first passage time of polymers with length $M=100$ against effective energy are plotted. Circle, square and diamond stand for $\lambda=1, 2$ and $\lambda=6$ respectively. Green color comes for the cases of $\alpha=0$ and red ones denote $\alpha=5$. Comparing two curves of $\lambda=6$ shows although at first ($EBE=-4$) the exponential distribution leads to a larger translocation time in smaller chaperones affinity ($EBE\geq-3$) it reverse. Inset is a zoom of the parent plot. The plots contain the error bar but it is small and not well visible.}
\label{TEn}
\end{figure}

Chaperone size is defined as $\lambda\sigma$ and as we said before, the $\lambda$ supposed to be an integer. We changed it from $1$ to $10$ in different EBEs and for distinct chaperones distributions. In order to clarify the important results of chaperones distribution from the sequence and focusing on the effects of energy, in contrast to our previous work \cite{AbdPLA16}, here we translocate homopolymers. The results is promising. Although in the uniform distribution of chaperones increasing the affinity of chaperones will always decrease translocation time of the polymer, it is not the case in exponential distribution and we have a minimum in the T versus EBE plot for chaperones size $\lambda>1$ (see figure \ref{TEn}). In spite of $\lambda=1$, in the case of $\lambda\geq2$ there is a contest between two important factors, affinity of the chaperone right near the wall and unbinding chance of chaperones from other sites. As we discussed in section \ref{theoexp} the competition will led to a minimum in T versus EBE curve (see figure \ref{TEn}). The competition is also important in calculating scaling exponent of the translocation time versus polymer length; $\beta$ ($T\sim M^\beta$).

\subsection{Scaling exponent; $\beta$}
\label{exp}
Scaling exponent of T versus M, $T\sim M^\beta$, is an important parameter to be calculated in polymer translocation literature. Limitation of our computational power do not allow us to simulate translocation of large polymers. This parameter helps people to predict translocation time of large polymers by knowing its trend in smaller lengths. It also helps to determine the translocation regime, whether it is diffusive or biassed.

\begin{figure}
\includegraphics[width=12cm]{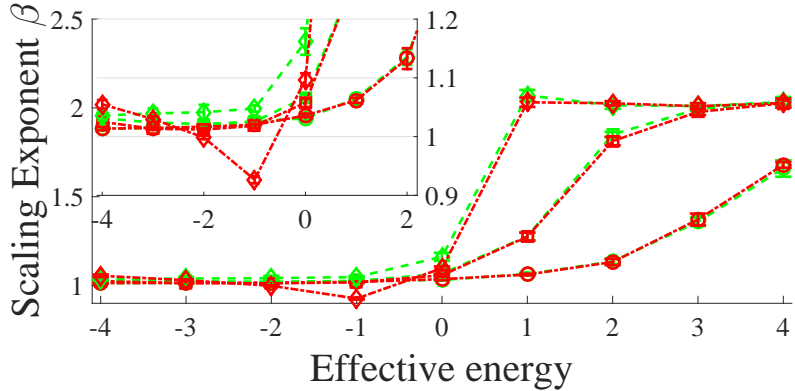}
\centering
\caption{Scaling exponent $\beta$ of the mean first passage time against length, $T\sim M^\beta$, is plotted with respect to $E_{eff}$ for different exponential rates of the chaperones distribution. The exponents are calculated by fitting the power function on the mean translocation time of the homopolymers with length of $100$ to $300$ monomers. In negative $E_{eff}$ we are in the ballistic regime with $\beta=1$, but in positive $E_{eff}$ the translocation regime of the polymer depends on the chaperones spatial distribution's exponential rate; $\alpha$.}
\label{Exponent}
\end{figure}

Scaling exponent $\beta$ is sketched versus EBE in figure \ref{Exponent} for different chaperone size and different distribution parameter, $\alpha$. The exponent is found by fitting power function on translocation time of the polymers with length between $50$ and $100$. As the inset shows in the case of $\alpha=5, \lambda=6$ we will reach to a minimum at $EBE=-1$. Generally this behavior is seen in cases of $\alpha\geq1$ and $\lambda\geq2$. The minimum point depends on the parameters $\alpha, \lambda$.

This behavior is due to competition between binding probability near the wall and it's unbinding from the polymer in further sites. As it has been shown in our previous works \cite{AbdPRE11}, scaling exponent is rely upon the lengths in which we choose to calculate the scaling exponent $\beta$. In larger polymers there is more time to see the chaperone's effect and increasing the polymer length will cause the exponent $\beta$ to decrease. Our simulation results, which has been done for polymer length between $50$ and $300$, indicate although the exponents depends on the length in which it has been calculated, the minimum location itself is length independent. Supper-diffusion, which is used in some cases as a sign of active transport, shows itself here, in some special cases, \textit{e.g.} the case of $\alpha=5, \lambda=6$ as shown in figure \ref{Exponent}. In appendix \ref{app3} using convection-diffusion equation and its dependency on P{\'e}clet number, we will discuss the possibility of supper-diffusion in more detail. The velocity dependency on polymer length in the limit of large P{\'e}clet number leads us to the supper-diffusion and taking its effect into account enable us to predict the minimum amount of scaling exponent $\beta$ (see equation \ref{betap} and its following discussion).

\subsection{Mean waiting time}
\label{mwt}
\begin{figure}
\includegraphics[width=10cm]{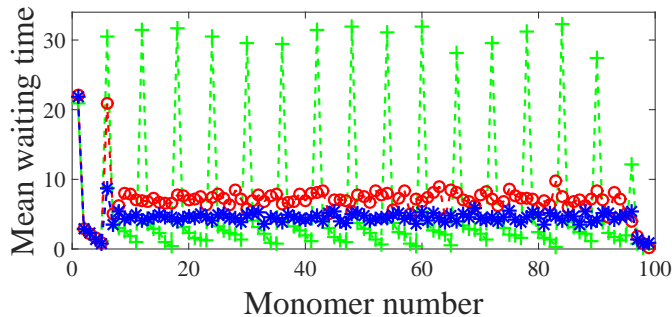}
\centering
\caption{Mean waiting time of polymers with length $M=100$ in vicinity of chaperones of size $\lambda=6$ and effective energy $E_{eff}=-4$ with three different spatial distribution sketched against monomer number. Green-plus calls for uniform distribution. Red-circle and blue-star stand for exponential spatial distribution of chaperones be rates $\alpha=5$ and $\alpha=10$, respectively. The peaks in uniform distribution are located in monomer numbers of integer multiple of $6$. Increasing the exponential rate $\alpha$ will decrease and diminish the peaks height.}
\label{mwtp4}
\end{figure}

Details of the translocation can be seen from its waiting times. Mean Waiting Time (MWT) of the translocation for chaperones of size $\lambda=6$ and $E_{eff}=-4$ for uniform chaperone's distribution and $2$ different rates of exponential distribution, $\alpha=5$ and $\alpha=10$, is shown in figure \ref{mwtp4}. After binding a chaperone near the wall due to its high affinity, polymer could not come back and soon will go to the right. It translocate to the right and left, until an empty space of the size $\lambda$ becomes available. Thus you see the peaks in monomer numbers of $6$, $12$, \ldots, which are integer multiples of $\lambda=6$ in uniform spatial distribution of chaperones. But in the case of exponential distribution, chaperone could unbind from the polymer and as a result the peak will be disappeared in large enough exponential rates, $\alpha$.

\begin{figure}
\includegraphics[width=10cm]{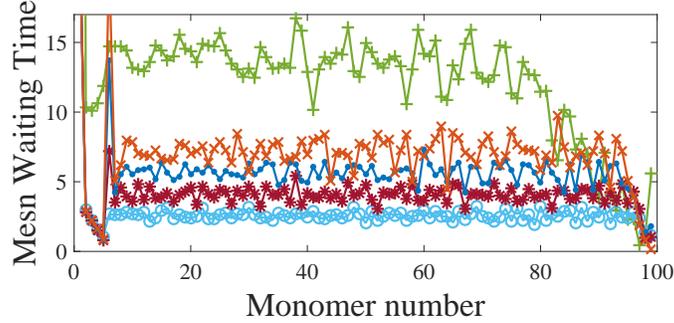}
\centering
\caption{Mean waiting time of polymers in vicinity of binding chaperones of size $\lambda=6$ and with exponential rate of $\alpha=5$ is plotted against monomer number for different chaperones affinities. Effective energies of $E_{eff}=0, -1, -2, -3$ and $E_{eff}=-4$ are shown by green-plus, blue-circle, red-star, magenta-dot and black-cross respectively. Due to their exponential rates $\alpha=5$ we do not see peaks as in figure \ref{mwtp4}. Increasing binding chaperones affinity will not always decrease the translocation time and the minimum of the waiting times are seen at $E_{eff}=-1$ not at $E_{eff}=-4$.}
\label{mwta5}
\end{figure}

As we discussed in previous section, in spite of the uniform distribution, in the case of exponential distribution of chaperones, increasing affinity ($E_{eff}$) will not always increase the translocation velocity. Figure \ref{mwta5} shows MWT of polymers in vicinity of chaperones with size $\lambda=6$ in which distributed exponentially by rate $\alpha=5$, with respect to monomer number for different affinity of $E_{eff}=-4,-3,-2,-1,0$. We see a minimum at $E_{eff}=-1$ which results from contest between binding just near the wall and unbinding in further sites. Correspondingly increasing the exponential rate will not consistently decrease the translocation time of the polymer.

\begin{figure}
\includegraphics[width=10cm]{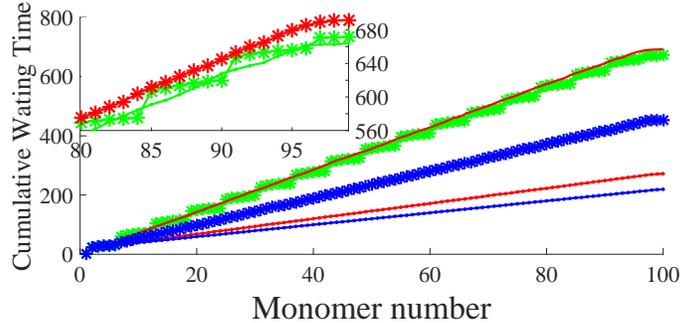}
\centering
\caption{Cumulative waiting time of polymers in vicinity of binding chaperones of size $\lambda=6$ and with different affinities $E_{eff}=1,4$ and in distinct spatial distribution is plotted against monomer number $m$ to be translocated. Green denotes uniform spatial distribution of chaperones. The colors yellow and blue stand for spatial exponential distribution of rates $\alpha=5$ and $\alpha=10$ respectively. We use dot for $E_{eff}=-1$ and star for $E_{eff}=-4$. The inset is a zoom of the parent figure to show the difference between $E_{eff}=0$ with uniform distribution and $E_{eff}=-4$ with exponential distribution of rates $\alpha=5, 10$.}
\label{CMT}
\end{figure}

As the result in inset of the figure \ref{CMT} shows increasing the chaperones exponential distribution rate, $\alpha$ from $0$ to $5$ will increase the translocation time rather than decrease it.

\section{Conclusions}
\label{conc}

Increasing affinity of the chaperones will increase their binding probability according to Boltzmann equation. Introducing spatial exponential distribution of chaperones in the problem makes the problem more complex. Now, binding (unbinding) probability will decrease (increase) by going away from the wall. Due to parking lot effect in the case of $\lambda>1$, this fact cause a contest between binding and unbinding. In large affinities the chaperones could not unbind and the minimum translocation time becomes $(\lambda+1)M$ (see \ref{lambda}) but in certain affinities chaperones will bind near the wall and unbind in other sites and as a results polymer can decrease its translocation time for larger chaperones $\lambda>1$ and make it possible to translocate even by $T_{min}=2M$ which is the minimum time for the case of $\lambda=1$. In this case the scaling exponent of translocation time versus length $\beta$ could be even as small as less than one!, which means \textit{e.g.} doubling the length will increase the time by less than itself. The results comes from finite size effect and in large enough length minimum of the scaling exponent $\beta$ come back to $\beta=1$.

\begin{appendix}

\section{Maximum and minimum of mean first passage time}
\label{app2}

\subsection{Maximum velocity}
\label{app21}

Here we suppose that the effective binding energies, $E_{eff}$, are negative and large. Chaperones will bind to the polymer as soon as they find an enough free space. Thus the time, $T_{min}$, equals to:

\begin{equation}
\label{t}
T_{min}=\frac{M}{\lambda}t_\lambda+t_\Delta\simeq\frac{M}{\lambda}t_\lambda, \Delta=M-\lambda[\frac{M}{\lambda}]<\lambda.
\end{equation}

where $t_\lambda$, is the time it takes for a bare polymer to translocate over a distance $\lambda\sigma$, $M$ is the total number of monomers and in large enough $M$ one could ignore $\Delta$. Forward and backward transition rates; $W^\pm(m)=\frac{1}{2\tau_0}$ for $1<m<\lambda$ and at the boundaries, reflective at first ($m=1$) and absorbing at the end ($m=\lambda+1$); $W^-(1)=W^\pm(\lambda+1)=0$ and $W^+(1)=\frac{1}{2\tau_0}$. Recognize that in our simulation the polymer may also remain in place when there is a chaperone near the wall and as a result the forward transition rate at the first boundary becomes $W^+(1)=\frac{1}{2\tau_0}$ rather than $\frac{1}{\tau_0}$.

Knowing the transition rates, one could calculate the mean first passage time in the corresponding boundary conditions, as follows (Ref.\cite{Ga}):

\begin{equation}
\label{MFPT}
T=\sum_{m=1}^\lambda
\left[\Phi(m)\sum_{m'=1}^m\frac{1}{W^+(m')\Phi(m')}\right],
\end{equation}
where we made use of the abbreviation
\begin{equation}
\Phi(m)=\prod_{u=2}^m\left[\frac{W^-(u)}{W^+(u)}\right].
\end{equation}

In this case the $\Phi(m)=1$ and thus the $T_{min}$ will becomes:

\begin{equation}
\label{lambda}
T_{min}=\frac{M}{\lambda}\left(\sum_{m=1}^\lambda
\left[2\times\sum_{m'=1}^m\frac{1}{1}\right]\right)=\frac{M}{\lambda}\left(2\times\sum_{m=1}^\lambda
\left[m\right]\right)=M(\lambda+1).
\end{equation}

Here one should pay attention that if we don't allow to the polymer to keep in palace at reflective boundary the transition rate $W^+(1)=1$ and the $T_{min}$ will becomes $M\lambda$.

\subsection{Minimum velocity}
\label{app22}

Following the above procedure and considering the fact that in this case the there isn't any chaperone bound on the polymer we will reach to the following results:

\begin{equation}
T_{max}=M(M+1).
\end{equation}

\section{Estimating the scaling exponent $\beta$}
\label{app3}

In a biased translocation one could write the continuum limit of the master equation (Eq.\ref{me}), the convection-diffusion equation, as\cite{AbdJCP11}:

\begin{equation}\label{condif}
  \frac{\partial\mathcal{P}(x,t)}{\partial t}+V\frac{\partial\mathcal{P}(x,t)}{
\partial x}=D\frac{\partial^2\mathcal{P}(x,t)}{\partial x^2},
\end{equation}

where $\mathcal{P}(x,t)$ denotes the probability density to find the chain at translocation coordinate $x$ at time
$t$. $V$ and $D$ are effective velocity and diffusion of the process. Solving the equation for the transmission mode, reflective at first and absorbing at the end, we will reach to the following equation for the mean translocation time:

\begin{equation}\label{cdt}
  T=\frac{L^2}{D}\left[\frac{1}{2Pe}-\frac{1}{4Pe^2}\Big(1-e^{-2Pe}\Big)\right],
\end{equation}

in which $L$ stands for polymer length and $Pe$ denotes the P{\'e}clet number; a dimensionless parameter comparing the
respective intensity of drift and diffusion \cite{redner}:

\begin{equation}
Pe\equiv\frac{VL}{2D}=\frac{1}{2}PM
\end{equation}\label{peclet}

in which $P$ denotes probability of near the wall binding site, to be occupied by a chaperone \cite{AbdJCP11}. Defining $\mu\equiv\frac{V}{D}$, scaling exponent $\beta$ can be write as:

\begin{equation}
\begin{aligned}
  \beta=&L\frac{\partial \ln(T)}{\partial L}\\
  =&L\frac{\partial}{\partial L}\ln(\frac{L}{D\mu}\left[1-\frac{1}{L\mu}\Big(1-e^{-L\mu}\Big)\right])\\
  =& 1-L\frac{\acute{\mu}}{\mu}+L\frac{\partial}{\partial L}\ln\left[1-\frac{1}{L\mu}\Big(1-e^{-L\mu}\Big)\right]
\end{aligned}\label{beta}
\end{equation}

in which $\acute{\mu}=\frac{\partial \mu}{\partial L}$. Thus:

\begin{equation}
\begin{aligned}
  \beta=& 1-L\frac{\acute{\mu}}{\mu}+\left[\frac{(1+L\frac{\acute{\mu}}{\mu})\Big(1-(1+L\mu)e^{-L\mu}\Big)}{L\mu-1+e^{-L\mu}}\right].
\end{aligned}
\end{equation}

For small values of P{\'e}clet number, diffusion-dominated regime, ($L\mu\rightarrow 0$):

\begin{equation}
\begin{aligned}
  \lim_{L\mu\rightarrow 0}\beta\approx& 1-L\frac{\acute{\mu}}{\mu}+(1+L\frac{\acute{\mu}}{\mu})\left(1-\frac{1}{3}L\mu+\frac{1}{18}(L\mu)^2+O((L\mu)^2)\right)\leq2,
\end{aligned}
\end{equation}

while for large P{\'e}clet number, drift-dominated regime, ($L\mu\rightarrow\infty$):

\begin{equation}\label{betap}
\begin{aligned}
  \lim_{L\mu\rightarrow\infty}\beta\approx& 1-L\frac{\acute{\mu}}{\mu}+\left(\frac{1+L\acute{\mu}/\mu}{L\mu}\right)
\end{aligned}
\end{equation}

which in the case of $\acute{\mu}>0$ will led to even scaling exponent of less than one, which is quite rare and usually unexpected.

Notice that in the limit of infinite polymers ($L\rightarrow\infty$), mean first passage time will be equal to $L/V$ which leads to:

\begin{equation}
\begin{aligned}
  \lim_{L\rightarrow\infty}\beta=L\frac{\partial \ln(L/V)}{\partial L}\approx& 1-L\frac{\acute{V}}{V}.
\end{aligned}
\end{equation}

where $\acute{V}=\frac{\partial V}{\partial L}$ and supposing the constant diffusion $D$ confirmed the limits of equation \ref{betap}.

\begin{figure}
\includegraphics[width=10cm]{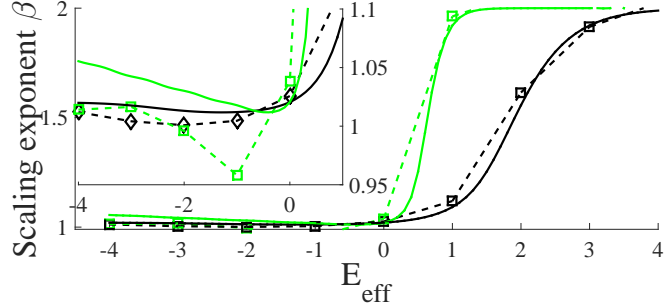}
\centering
\caption{Scaling exponent $\beta$ with respect to effective energy of chaperones distributed exponentially of rate $\alpha=5$ and with two different size of $\lambda=2, 6$ are compared with theory calculated from estimated $P_{nw}$ in equation \ref{Pnwapp}.}
\label{ThoSim}
\end{figure}

Supposing $\mu$ to be independent of length one can easily calculate the scaling exponent $\beta$ from equation \ref{beta} and by substituting $P$ in P{\'e}clet number(equation\ref{peclet}) from equation \ref{Pnwapp}. This theoretical result is compared by simulation in figure \ref{ThoSim}. Although this theory predicts the scaling exponents behavior well, the scaling exponents are always greater than one and it cannot predict the minimum of $\beta$ correctly. Interestingly by taking into account the discussion here and calculating $\frac{\partial V}{\partial L}$ from simulation result for the case of $\lambda=6, \alpha=5$ and in its minimum, $E_{eff}=-1$, we will reach to the result of $\beta=0.97$ which is within the error bar of the exact value from the simulation.

\end{appendix}


\bibliographystyle{elsarticle-harv}
\bibliography{MyReferences}

\end{document}